# Sustainable photocatalytic $CO_2$ conversion using microalgae as a carbon-negative scavenger

Ho Truong Nam Hai, https://orcid.org/0000-0001-8804-6166 [ab] Augusto Ducati Luchessi https://orcid.org/0000-0003-2080-3524 [c] Makoto Arita, https://orcid.org/0000-0001-6785-3890 [d] Qixin Guo https://orcid.org/0000-0002-9471-2353 [e] and Kaveh Edalati https://orcid.org/0000-0003-3885-2121 *[ab]

Photocatalytic $CO_2$ conversion driven by solar energy is a highly promising approach in addressing rising atmospheric $CO_2$ levels; however, its practical application remains limited by low conversion efficiency. In this study, a new strategy to enhance $CO_2$ reduction toward CO and $CH_4$ is proposed through the employment of microalgae as a sacrificial agent, and the efficiency is compared with conventional $CO_2$ conversion without and with the use of microplastics as sacrificial agents. To realize this strategy, an $AB_2O_6$-type high-entropy oxide (HEO), $(Cs_{1/7}Ba_{4/7}Bi_{2/7})(Nb_{1/2}Ta_{1/2})_2O_6$, with bi-polymorphy of layered perovskite and pyrochlore, is rationally designed. The HEO incorporates alkali metal cesium and alkaline earth metal barium to increase surface basicity for $CO_2$ chemisorption, bismuth with its stereochemically active lone pairs for localized polarization and charge separation, and tantalum and niobium to form octahedral crystalline frameworks for charge transport. The utilization of microalgae during photocatalytic reactions leads to a remarkable enhancement in $CO_2$ conversion compared to catalysis with or without using microplastics, with CO and $CH_4$ production increasing by 10- and 4-fold, respectively, compared to the system using only HEO. These findings not only demonstrate a new family of polymorphic $AB_2$-type HEOs for photocatalysis but also show the potential of microalgae as a sustainable sacrificial agent, offering an environmentally friendly pathway for efficient $CO_2$ capture (through photosynthesis by microalgae) and $CO_2$ conversion (through photocatalysis by HEOs).

## Introduction

As a central component of the carbon cycle, the $CO_2$ concentration in a stable range is critical to sustaining life and preserving ecological balance on the Earth.[1] However, the excessive exploitation and use of fossil fuels, along with human-induced environmental issues such as deforestation and other industrial activities, have led to the release of large amounts of $CO_2$ into the atmosphere.[2] In addition to green energy transition policies to replace fossil fuels, methods for converting $CO_2$ into fuel products such as CO, $CH_4$, or other chemical compounds have emerged as a strategic approach for carbon mitigation and energy exploitation. Photocatalysis is a promising and sustainable method for $CO_2$ conversion,[3,4] relying on suitable catalysts and solar radiation as the energy source. Nevertheless, properties of photocatalysts, including band structure, surface features, active sites, charge recombination under irradiation, stability, and toxicity, significantly affect the efficiency of the process and its practical application in $CO_2$ reduction.[5] In addition, most photocatalytic $CO_2$ conversion reactions are conducted in aqueous environments.[5] As a result, photogenerated electrons must compete between $CO_2$ reduction and $H_2O$ reduction, posing a major challenge in the rational design of effective catalysts. In addition to challenges with the design of the catalyst that should be addressed, the need for a sacrificial agent is another issue in $CO_2$ conversion because most sacrificial agents, like alcohol, are produced by chemical methods that involve $CO_2$ emissions.

To address the first issue, recent research has increasingly focused on the photocatalytic activity of Nb- and Ta-based materials.[6–9] Niobium and tantalum are group V transition metals in the periodic table and possess relatively low electronegativity. In their oxidized states, they typically exhibit a $d^0$ electronic configuration.[10] The presence of empty d orbitals can facilitate efficient charge transfer.[11] Moreover, niobium and tantalum can provide octahedral frameworks to accommodate other elements. For example, several alkali and alkaline earth metal oxides, such as $KNbO_3$,[12,13] $NaNbO_3$,[13] and $BaNb_2O_6$,[14] have been shown to be environmental semiconductors with high potential for photocatalytic applications. These Nb- and Ta-based catalysts show strong oxidative power while maintaining excellent structural stability.[15–17]

[a.] *WPI, International Institute for Carbon Neutral Energy Research (WPI-I2CNER), Kyushu University, Fukuoka 819-0395, Japan.*
[b.] *Department of Automotive Science, Graduate School of Integrated Frontier Sciences, Kyushu University, Fukuoka 819-0395, Japan.*
[c.] *Laboratório de Biotecnologia BraPhyto, Faculdade de Ciências Aplicadas (FCA), Universidade Estadual de Campinas (UNICAMP), 13484-350, Limeira, SP, Brazil*
[d.] *Department of Materials, Kyushu University, Fukuoka 819-0395, Japan*
[e.] *Department of Electrical and Electronic Engineering, Synchrotron Light Application Center, Saga University, Saga 840-8502, Japan*
† Corresponding author (E-mail: kaveh.edalati@kyudai.jp; Phone: +81 92 802 6744).

To further enhance photocatalytic efficiency and structural robustness, high-entropy ceramics have emerged as a novel design concept by integrating multiple cations into a single lattice.[18] The use of niobium and tantalum to generate octahedral frameworks of high-entropy oxides (HEO) offers opportunities to synergistically modulate electronic structure and catalytic properties by incorporating various cations. However, most attempts to use niobium and tantalum result in chemical separation of elements.[18-21] One possible strategy to use the benefits of niobium and tantalum is designing stable crystal structures similar to perovskites ($ABO_3$) or layered perovskites ($AB_2O_6$), where the A site is an element with a large atomic radius, such as alkali, alkaline earth, or rare earth elements, while the B site is an element with a smaller radius, such as transition metals. While there are still significant limits in selecting the elements for the $ABO_3$-type perovskites, as discussed elsewhere,[20] the $AB_2O_6$-type compositions are expected to incorporate a wider range of elements in A and B sides with a wider possible polymorphism.

Regarding the second issue, designing photocatalytic processes that simultaneously use both electrons and holes is crucial for enhancing $CO_2$ reduction efficiency and improving product yields. To consume holes, hole scavengers such as alcohol are commonly introduced into photocatalytic systems. However, the use of alcohol significantly increases the cost of photocatalytic $CO_2$ conversion compared with alternative approaches such as biological, thermochemical, or electrochemical conversion.[5] Moreover, production of alcohol involves $CO_2$ emissions, which negatively affect the carbon-neutrality of photocatalysis. Therefore, it is necessary to replace alcohol with other sacrificial agents that are inexpensive, abundant, and nature-friendly. Based on these requirements, microalgae can be considered a candidate for a sacrificial agent. Microalgae are well-known as photosynthetic organisms that are widely distributed in most aquatic environments.[22] Under conditions from normal to extremely harsh environments of light intensity, pH, and nutrient availability, microalgae can still sustain growth.[23] Unlike alcohol, whose production emits $CO_2$, cultivating microalgae captures $CO_2$ from the atmosphere through photosynthesis. The current study hypothesizes that the harvested microalgae can subsequently contribute to $CO_2$ conversion by being employed as a sacrificial agent in photocatalysis.

In this study, a photocatalytic process is introduced in which microalgae are employed as a sacrificial agent to enhance $CO_2$ conversion. A novel high-entropy material with the $AB_2O_6$ structure is designed and synthesized by high-pressure torsion (HPT) [24] and subsequent calcination for this process. The A site contains alkali metal cesium and alkaline earth metal barium to enhance $CO_2$ chemisorption and bismuth to stereochemically activate lone pairs for charge separation via localized polarization.[11] In addition, a higher atomic percentage of barium compared to bismuth and cesium was chosen to (i) ensure proper occupation of the A-site cations within the cuboctahedral cavities, and (ii) limit the volatility of cesium during the high-temperature synthesis. Niobium and tantalum occupy the B site, constructing a robust octahedral framework that promotes charge carrier transport. Analysis using a catalytic system, illustrated in Fig. 1, confirms that the integration of microalgae for $CO_2$ conversion in the presence of the HEO $(Cs_{1/7}Ba_{4/7}Bi_{2/7})(Nb_{1/2}Ta_{1/2})_2O_6$ leads to higher CO and $CH_4$ production, compared to conventional photocatalysis or photocatalysis utilizing polyethylene terephthalate (PET) microplastics as a sacrificial agent. These results offer a new sustainable route to both $CO_2$ capture (through photosynthesis) and $CO_2$ conversion (through photocatalysis). A main benefit of this route is that its scavenger does not produce $CO_2$ during its production (unlike alcohol), but it captures and converts $CO_2$, which can be considered a carbon-negative sacrificial agent.

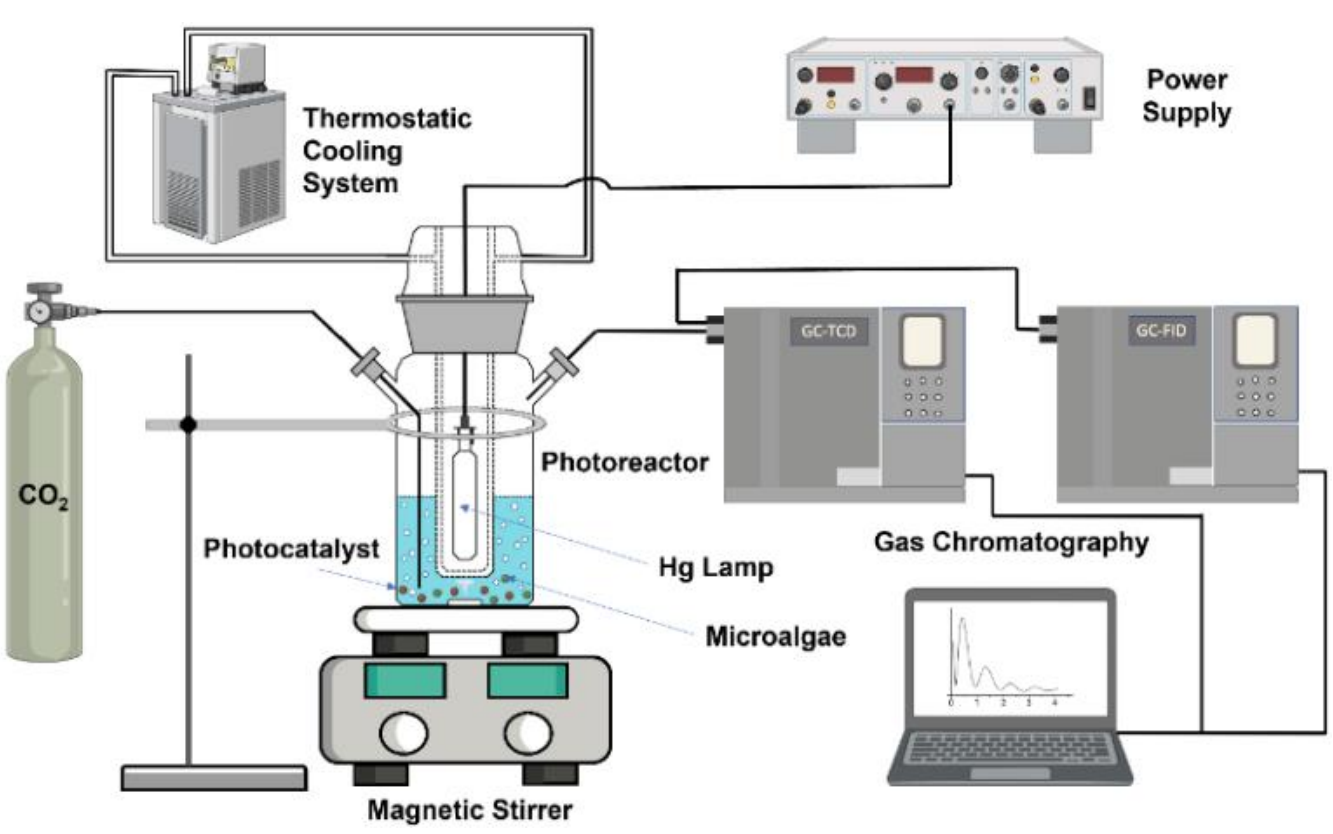

Fig. 1. Schematics of the $CO_2$ conversion experiment using microalgae as sacrificial agent.

## Materials and methods

### Reagents

Five metal oxides, including cesium tantalum oxide ($CsTaO_3$, 99.9 %), barium peroxide ($BaO_2$, 99 %), bismuth trioxide ($Bi_2O_3$, 99.9 %), niobium pentoxide ($Nb_2O_5$, 99 %), and tantalum pentoxide ($Ta_2O_5$, 99.9 %), were obtained from Mitsuwa Pure Chemical, Kojundo, and Fujifilm, Japan. Sodium bicarbonate ($NaHCO_3$) and PET plastic powder were purchased from Nacalai Tesque and Fujifilm, Japan, respectively.

### Synthesis of photocatalyst

To synthesize the HEO, the initial five metal oxides were calculated and weighed to form the formula $(Cs_{1/7}Ba_{4/7}Bi_{2/7})(Nb_{1/2}Ta_{1/2})_2O_6$. Since some losses of Cs can occur at high temperatures, the amount of the element was calculated over the nominal values to reach the final composition after the synthesis. By considering $BaNb_2O_6$ as a reference, barium in the A site was kept high, while Cs was kept low to avoid chemical separation in the form of $CsTaO_3$. These oxides were mixed by grinding in ethanol for 30 min. The resulting powder was then compressed into a small disc with a volume of 0.063 $cm^3$ and a diameter of 1 cm. Subsequently, the disc was processed by HPT [24] under a pressure of 6 GPa, with 3 rotations at a speed of 1 rpm at ambient temperature. The HPT process was repeated once more after grinding under similar HPT treatment. The sample was calcined at 1373 K for 24 h, reprocessed by HPT, and recalcined under identical conditions to ensure elemental homogeneity.

### Characterization of photocatalyst

The properties of the HEO were investigated by various analytical techniques. The crystal structure and phase composition of the HEO were determined through X-ray diffraction (XRD) using a Cu K$\alpha$ radiation source. Raman spectroscopy analysis with a laser light source at a wavelength of 532 nm was used to determine molecular vibrations in the HEO. Scanning electron microscope (SEM) and energy-dispersive X-ray spectroscopy (EDS) were used to confirm the morphology and elemental distribution at the microscale. Furthermore, these results were corroborated by transmission electron microscopy (TEM) and scanning transmission electron microscopy (STEM) analyses at the nanoscale. TEM and STEM micrographs were obtained at 200 keV, including bright-field (BF), dark-field (DF), selected area electron diffraction (SAED), and high-resolution analyses. Using synchrotron radiation (beamline BL11 at the Saga Light Source), X-ray absorption spectroscopy (XAS) was conducted, which provided high-intensity and tunable beams that enabled detailed analysis of the atomic structure of the material. XAS consists of two complementary regions, including X-ray absorption near-edge structure (XANES) and extended X-ray absorption fine structure (EXAFS). XANES is sensitive to the electronic structure and oxidation states, whereas EXAFS characterizes bonding configurations, structural disorder, and neighboring atomic arrangements in the HEO. X-ray photoelectron spectroscopy (XPS) with Al K$\alpha$ radiation and ultraviolet photoelectron spectroscopy (UPS) using a He I source were applied to identify elemental oxidation states and to estimate the valence band maximum (VBM) of the catalyst. UV-Vis spectroscopy in the 200-800 nm range was conducted to assess light absorption properties. Based on Kubelka-Munk theory and UV-Vis results, the band gap of the catalyst was determined. Charge carrier recombination involving photogenerated electrons and holes was investigated by the photoluminescence technique using a 325 nm excitation laser. In addition, oxygen vacancies in the samples were investigated by electron spin resonance (ESR) measurements performed at a microwave frequency of 9.4688 GHz to probe unpaired electrons.

### Microalgae cultivation

In this study, microalgae were employed as sacrificial agents in the photocatalytic $CO_2$ conversion. To collect microalgae, a laboratory-scale cultivation procedure was carried out. Initially, the wild-type *Chlamydomonas reinhardtii* CC-124 strain was supplied from the Chlamydomonas Resource Center (CRC), USA. The microalgae were cultured in petri dishes containing tris–acetate–phosphate (TAP) and agar medium, as suggested in the recipe on the CRC website. Specifically, 1 L of TAP medium was prepared from 20 mL of 1 M tris base, 1 mL of phosphate buffer II (10.8 g $K_2HPO_4$ and 5.6 g $KH_2PO_4$ per 100 mL), 10 mL of solution made from 20 g $NH_4Cl$, 5 g $MgSO_4.7H_2O$, and 2.5 g $CaCl_2.2H_2O$ per 500 mL, 1 mL of Hutner's trace elements, and 1 mL of glacial acetic acid. For a solid medium, 15 g $L^{-1}$ of agar was added. A continuous light source with an intensity of 0.048 mmol photons $m^{-2}s^{-1}$ was maintained throughout the cultivation period (a 24 h/day light cycle) at 293 K. For scaling up the amount, the cultures were transferred from petri dishes into Erlenmeyer flasks containing TAP medium under ambient temperature conditions. The flasks were agitated on a rotary shaker at 150 rpm to facilitate gas exchange for microalgal growth. Once the microalgae reached 110 mL, the culture was transferred to a 15 L large-scale reactor containing TAP medium. The culture was further stirred and aerated under atmospheric conditions, then filtered through a 220 nm membrane filter. After achieving the desired biomass, 2.0 g of microalgae were collected and centrifuged for 10 min. The harvested cells were subsequently dehydrated by lyophilization to obtain the final product. Finally, the samples were refrigerated at 253 K, transported, and preserved for further use. Prior to photocatalytic experiments, the microalgae were characterized using Fourier transform infrared spectroscopy (FTIR) coupled with single-reflection attenuated total reflectance (ATR).

### $CO_2$ conversion experiments

Fig. 1 illustrates the schematic of the $CO_2$ conversion experiment. An 856 mL cylindrical-shaped photoreactor was placed on a magnetic stirrer operating at a speed of 420 rpm. For sample preparation, 100 mg of catalyst was added to the reaction vessel, followed by adding 4.2 g of bicarbonate and 500 mL of deionized (DI) water as a neutralized aqueous environment using a magnetic stirrer. Subsequently, 100 mg of the PET plastic powder or microalgae cluster, as sacrificial agents for holes, were introduced, depending on the experiment. For comparison with a conventional sacrificial agent, a similar experiment was conducted using 450 mL of DI water and 50 mL of methanol, corresponding to 10 vol% of the total solution. The lid is then sealed to ensure an anaerobic environment for the experiment. The lid was equipped with two valves connected to a thermostatic cooling system, enabling the experimental temperature to be maintained at 291 K. Next, a mercury lamp (1.4 W $cm^{-2}$) was placed in the void inside the lid as a radiation source for the experiment. A continuous stream of $CO_2$ gas at a flow rate of 30 mL/min was then bubbled into the solution through an inlet pipe. Two gas chromatographs (GC), one connected with a methanizer and flame ionization detector (FID) and the other with a thermal conductivity detector (TCD), were connected to the outlet pipe to quantify gaseous products evolution ($H_2$, $CH_4$, and CO) generated from the photocatalytic $CO_2$ conversion. A control test was first conducted under off-light conditions for 30 min to make sure that $CO_2$ conversion did not occur in the absence of irradiation. Subsequently, the mercury lamp was switched on to initiate the reaction. Gas sampling was performed automatically at hourly intervals, and the concentration analysis data were displayed using monitoring software.

Following the photocatalysis, the reaction mixture was passed through a polytetrafluoroethylene wheel filter (Code: PTFE013045, pore size: 0.45 µm, radius: 6.5 mm) to separate the liquid phase. An aliquot of 20 mL of the resulting filtrate was placed in a separatory funnel and mixed with 5 mL of hexane, 2 mL of saturated KCl solution, and 2 mL of deionized water, then subjected to liquid–liquid extraction for 30 min. The upper organic layer was collected, and the lower layer was subjected to a second time extraction with an additional 5 mL of hexane to maximize the recovery of organic composition from the liquid phase. The organic extracts were subsequently dried by argon gas to concentrate the sample. For chemical composition identification, 1 μL of the concentrated extract was injected directly into gas chromatography-mass spectrometry (GC-MS). The GC analysis employed a CP-Sil 8CB capillary column with a split ratio of 50:1. The oven temperature was held at 313 K for 2 min, heated to 513 K at 4 K $min^{-1}$, and kept at 513 K for 2 min. Mass spectrometric analysis was carried out with an ion source temperature of 473 K, an interface temperature of 523 K, and an m/z (ratio of mass to charge) scanning range of 50-700.

## Results

### Crystal structure and composition

Fig. 2a describes the crystal structure of the material at different stages of synthesis. The XRD profile shows the transformation of the crystal structure from the initial mixture of oxides to new phases. It can be observed that after the first processing stage, where the sample was treated by HPT twice and subsequently calcinated, the resulting material (HPT + HPT + C) exhibits new diffraction peaks that differ in position and intensity compared to the peaks of the initial oxide mixture. This indicates a rearrangement of the elements in the crystal lattice structure to form a new oxide. This arrangement is further enhanced by repeating the HPT process and calcination. In this second processing stage, the peaks associated with the binary oxides completely vanish, while the new peaks become more separated and their intensity becomes higher. Fig. 2b shows a more detailed observation of the HEO material after the synthesis process (HPT + HPT + C + HPT + HPT + C). The Rietveld refinement ($R_{wp}$ = 8.51%, $R_p$ = 6.07%, $R_e$ = 2.40%, $s$ = 3.54) indicates that the material contains two phases: 56 wt% cubic ($Fd\bar{3}m$ space group, $a = b = c$ = 10.5183 Å; $\alpha = \beta = \gamma$ = 90°) and 44 wt% tetragonal ($P4bm$ space group, $a = b$ = 12.519 Å, $c$ = 3.948 Å; $\alpha = \beta = \gamma$ = 90°). The $Fd\bar{3}m$ space group is associated with a pyrochlore-type structure of $AB_2O_6$ oxides.[25] Meanwhile, the $Pb4m$ space group is related to the $AB_2O_6$-type layered hexagonal perovskite structure, which is also similar to tetragonal tungsten bronze structures in tantalates, such as $SrTa_2O_6$ (unit cell: $a = b$ = 12.41 Å, $c$ = 3.90 Å, $\alpha = \beta = \gamma$ = 90°) or $BaTa_2O_6$ (unit cell: $a = b$ = 12.60 Å, $c$ = 3.95 Å, $\alpha = \beta = \gamma$ = 90°).[26] Fig. 2c depicts the Raman spectra of the HEO acquired at four randomly selected positions. All Raman spectra show no significant differences, indicating a high level of material homogeneity. In addition, the structure of the HEO is also better understood through the different vibrational modes corresponding to each Raman peak. First, a peak at 152 $cm^{-1}$ is related to lattice vibration.[4] Second, the peak at 252 $cm^{-1}$ is assigned to Ba–O vibration in the crystal structure, representing a feature similar to the A-site position in perovskites.[20] Third, the peak at 499 $cm^{-1}$ is assigned the $A_{1g}$ + $B_{1g}$ vibration.[27,28] Meanwhile, the peak at 616 $cm^{-1}$ can be attributed to the $A_{1g}$ vibration mode.[27,28] Finally, a low-intensity peak at 872.7 $cm^{-1}$ should be related to the vacancy defects.[20]

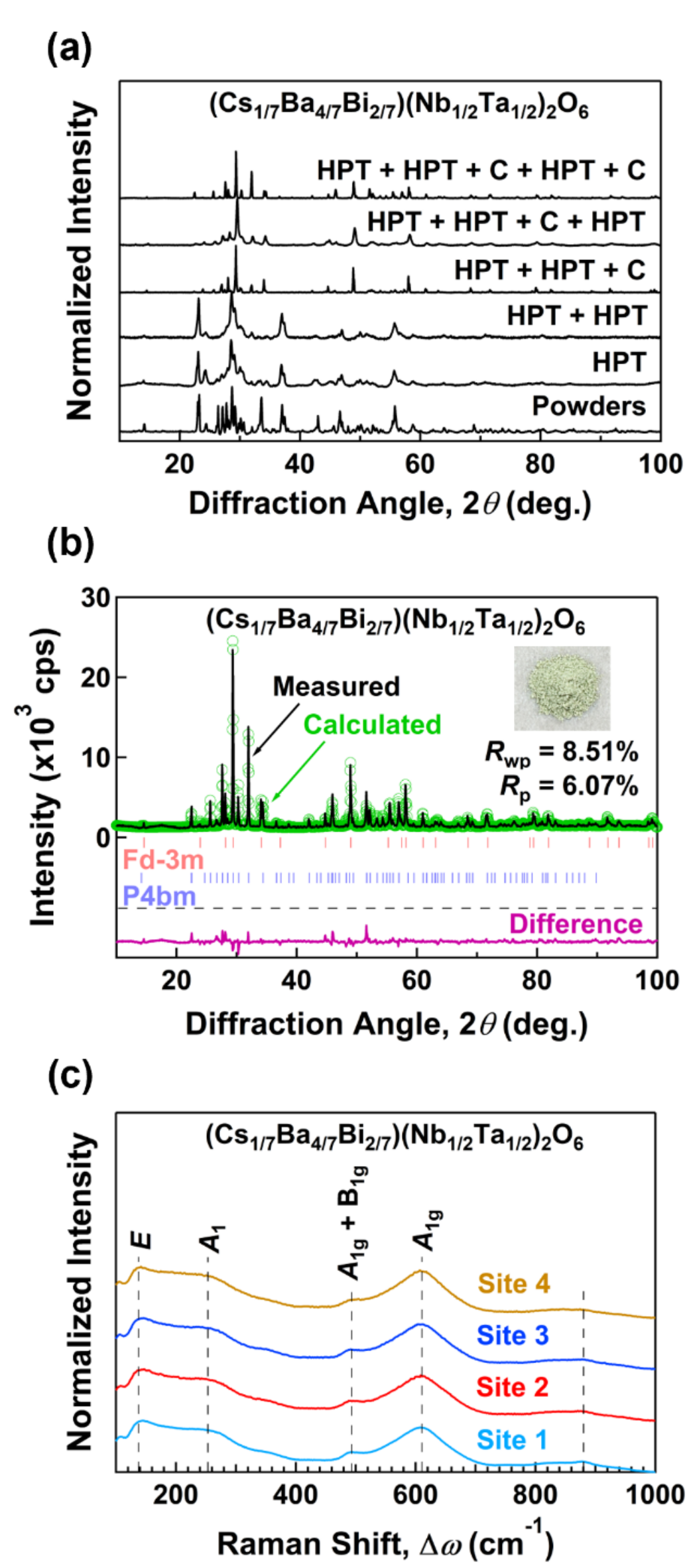

Fig. 2. Thermomechanical synthesis of HEO. (a) XRD spectra at each synthesis stage implemented by HPT and calcination, (b) XRD spectrum, corresponding Rietveld refinement, and real image (insert) of HEO. (c) Raman spectrum of HEO examined at four different sites.

Fig. 3 illustrates the elemental distribution in the HEO using two techniques: SEM-EDS and STEM-EDS. Homogeneous elemental distribution is evident at the micrometer scale in Fig. 3a and at the nanometer scale in Fig. 3b. Elemental atomic percentages measured by SEM-EDS and STEM-EDS are listed in Table 1, indicating that, within the precision limits of EDS for quantitative chemical analyses, the final composition reaches the designed one,

$(Cs_{1/7}Ba_{4/7}Bi_{2/7})(Nb_{1/2}Ta_{1/2})_2O_6$, after evaporation of some cesium [29] and bismuth.[30]

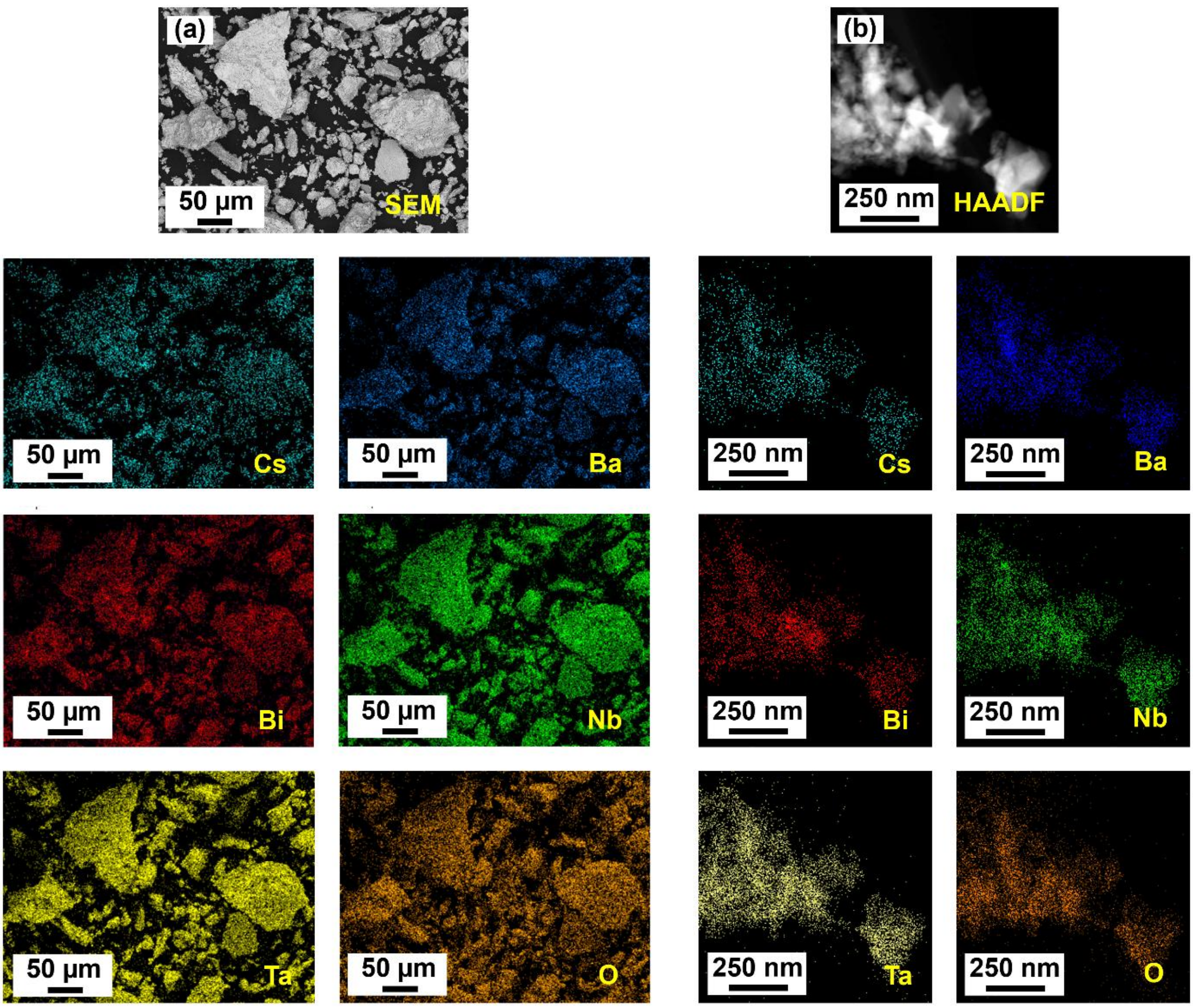

Fig. 3. Homogeneous arrangements of elements in HEO. (a) SEM micrograph with EDS mappings and (b) HAADF micrograph by STEM with corresponding EDS mappings for HEO.

Table 1. Elemental composition of HEO analyzed by SEM-EDS and STEM-EDS.

| | Composition (at%) | | | | | |
|---|---|---|---|---|---|---|
| | **Cs** | **Ba** | **Bi** | **Nb** | **Ta** | **O** |
| **SEM-EDS** | 1.4 | 6.1 | 2.5 | 9.0 | 8.7 | 72.3 |
| **STEM-EDS** | 1.4 | 6.4 | 3.1 | 9.9 | 7.6 | 71.6 |

The cation oxidation states and the oxygen reduction state in the HEO were determined by fitted XPS spectra, as presented in Fig. 4. The observed peaks for Cs $3d^{5/2}$, Ba $3d^{5/2}$, Bi $4f^{7/2}$, Nb $3d^{5/2}$, and Ta $4f^{7/2}$ appear at 724.1, 780.1, 159.1, 206.8, and 25.5 eV, respectively. Peak deconvolution of XPS spectra confirms the fully oxidized states of the cations in the HEO ($Cs^{1+}$, $Ba^{2+}$, $Bi^{3+}$, $Nb^{5+}$, and $Ta^{5+}$). The XPS spectrum and peak deconvolution of oxygen in the HEO reveal the presence of three peaks. Among them, the peak at 529.8 eV is assigned to lattice O 1s,[11] the peak at 531.5 eV is associated with hydroxyl groups and oxygen vacancies,[20] and the peak at 533.3 eV corresponds to adsorbed water on the HEO surface.[31]

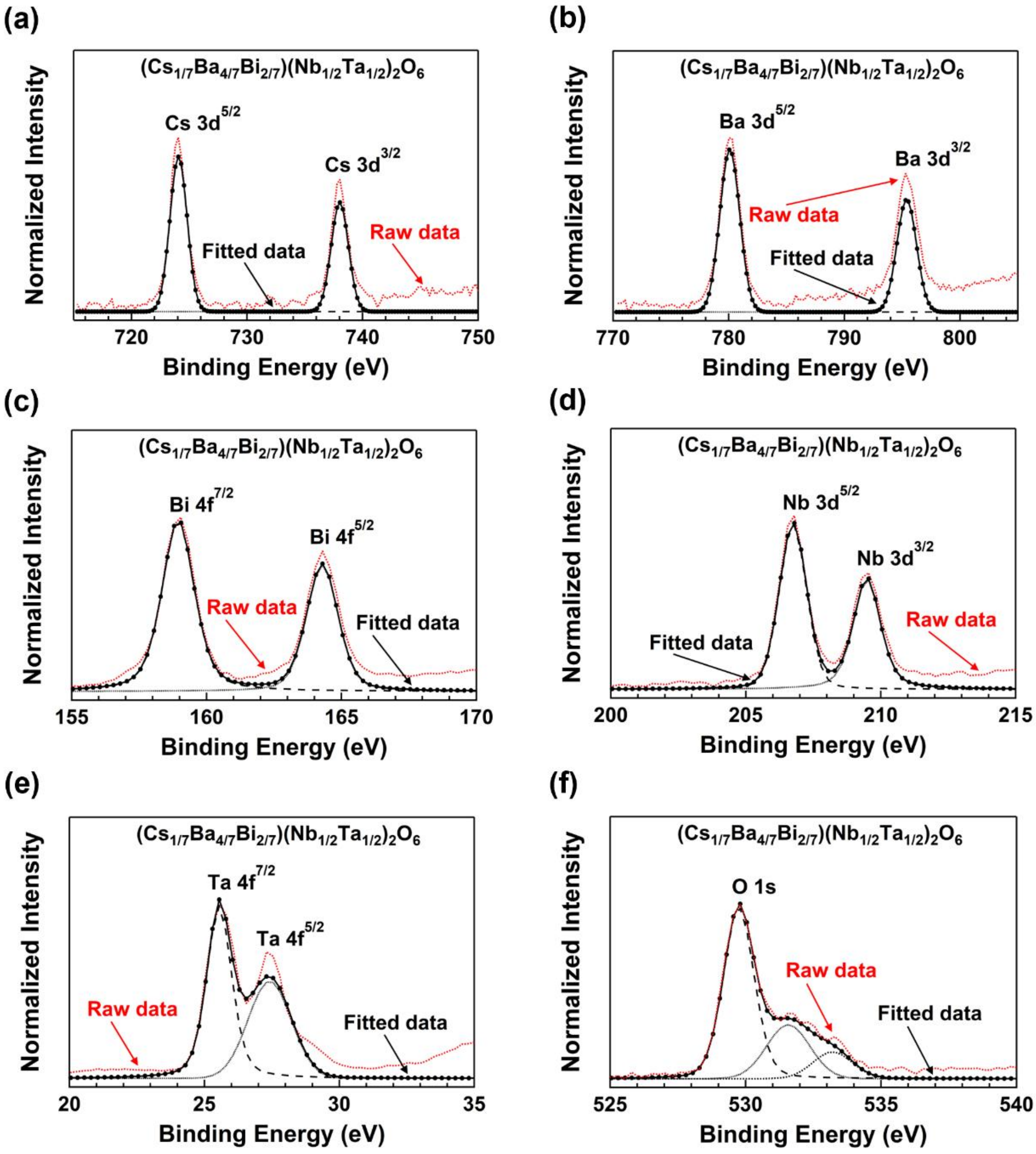

Fig. 4. Chemical state of metals and oxygen in HEO. XPS spectra for (a) cesium, (b) barium, (c) bismuth, (d) niobium, (e) tantalum, and (f) oxygen in HEO.

**Nanostructure and atomic structure**

Fig. 5 illustrates the nanostructural features of the HEO. The TEM (a) BF and (b) DF micrographs show the presence of nanograins in the HEO. These grains appear as a typical feature of the HPT process, which apparently remains even after high-temperature calcination.[24] Fig. 5c depicts the SAED pattern of HEO and shows concentric rings representing specific crystal planes of the cubic and tetragonal phases in the HEO. In addition, Fig. 5d and 5e also confirm the existence of cubic and tetragonal phases through HR micrographs and fast Fourier transforms, respectively. The observation of the nanostructure shows that, in addition to high-angle grain boundaries, dislocation cells are also present, as described in Fig 5f. Several studies have shown that the presence of dislocations and their pair vacancies can contribute to increased charge separation, and thus improve the efficiency of the photocatalysis process.[32,33]

The atomic structure characteristics of the four metals (Cs $L_3$-edge, Ba $L_3$-edge, Bi $L_3$-edge, Ta $L_3$-edge) in the HEO and their corresponding binary and ternary oxides are described using XANES and EXAFS spectra, as shown in Fig. 6. Based on the XANES spectra (Fig 6a, 6c, 6e, 6g), a similarity can be observed between the absorption edges of the metals in the HEO and the binary and ternary oxides, reflecting that the oxidation state and local electronic structure of the metals do not significantly differ. In addition, XANES provides further information about the oxidation states of the metals through the white line.[34] The white line is known as the strong, sharp absorption peak that appears right above the absorption edge. A higher white line reflects a higher oxidation state based on a higher 5d unoccupied state.[34] Therefore, it can be seen that barium and bismuth in the HEO have more unoccupied states compared to barium in $BaO_2$ and bismuth in $Bi_2O_3$. In contrast, the number of unoccupied states of cesium and tantalum in the HEO is lower than that of cesium in $CsTaO_3$ and tantalum in $Ta_2O_5$. Additionally, the region adjacent to the white line shows a shift toward lower energy levels for the Cs and Ba peaks, while the Bi peak shifts toward higher energy, and Ta remains unchanged. This indicates that the bond lengths of Cs–O and Ba–O in the HEO are shorter than in their corresponding binary oxides, whereas the Bi–O bond length is longer. It should be noted that the bond lengths achieved by binary oxides in this study are consistent with the literature. [35,36] Fig. 6b, 6d, 6f, and 6h provide information about the vicinity of the cations through the oscillations of the 50-1000 eV energy absorption spectrum on the absorption edge determined by EXAFS analysis.[37,38] In these figures, the high-intensity peak at approximately 1-2 Å typically characterizes oxygen bonded to the metal, while the subsequent peaks (second or third) usually describe the bonding of the metal to itself or other metals within a material. It can be observed that there are differences in most peak positions, except for one Ta peak at 1.5 Å. This confirms the preservation of the bonding between $Ta^{5+}$ and six $O^{2-}$ atoms in a $TaO_6$ octahedron. Meanwhile, the arrangement of A-site elements such as Ba, Bi, and Cs is completely different between the HEO and their corresponding binary oxides. These atomic structure characteristics also confirm the perovskite structure of HEO. Furthermore, these features can influence photocatalytic processes such as the conversion of $CO_2$ into $CH_4$ or CO products.

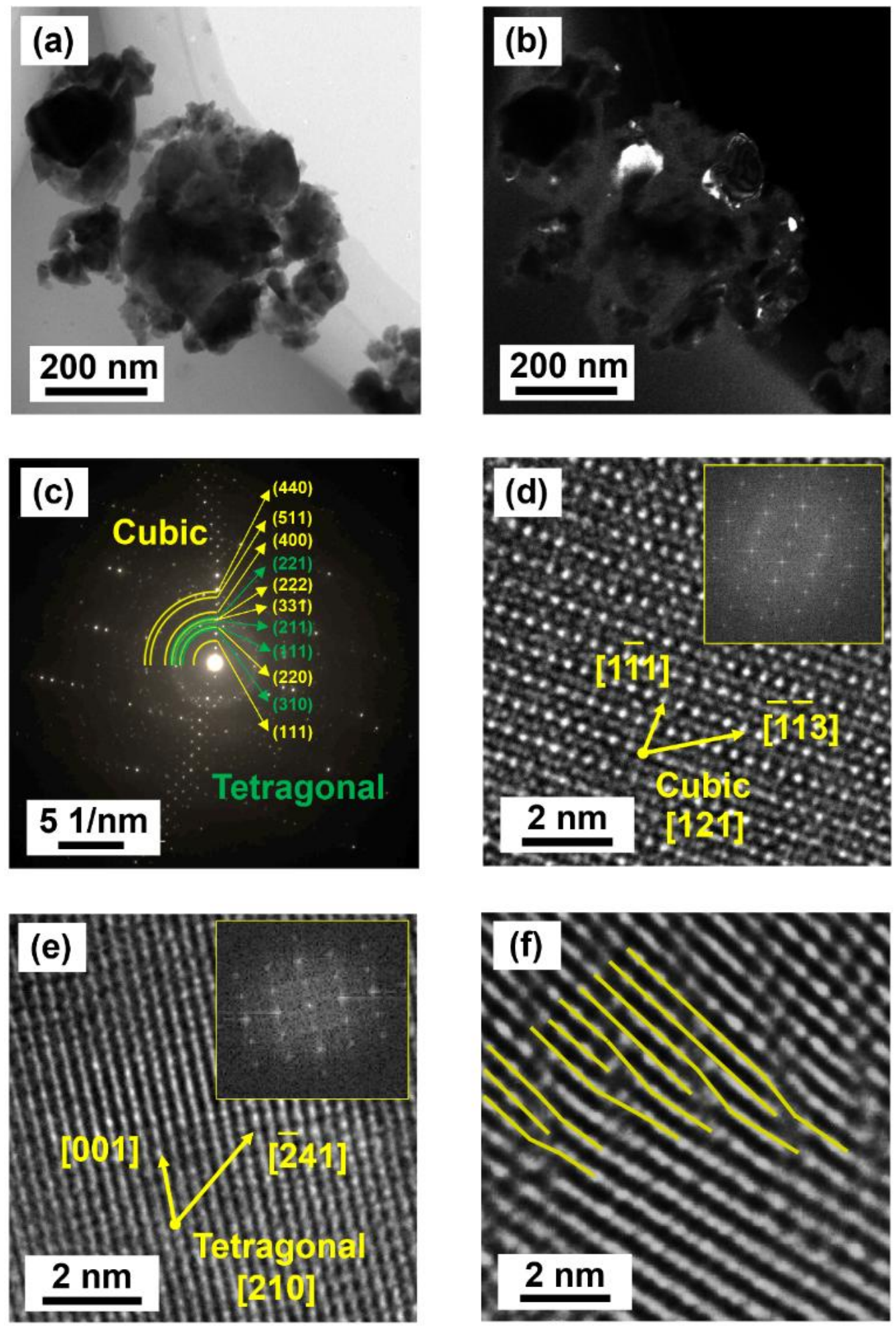

Fig. 5. Formation of nanograins and dislocation cells in HEO. (a) BF, (b) DF, (c) SAED, and (d-f) HR analyses of HEO taken by TEM.

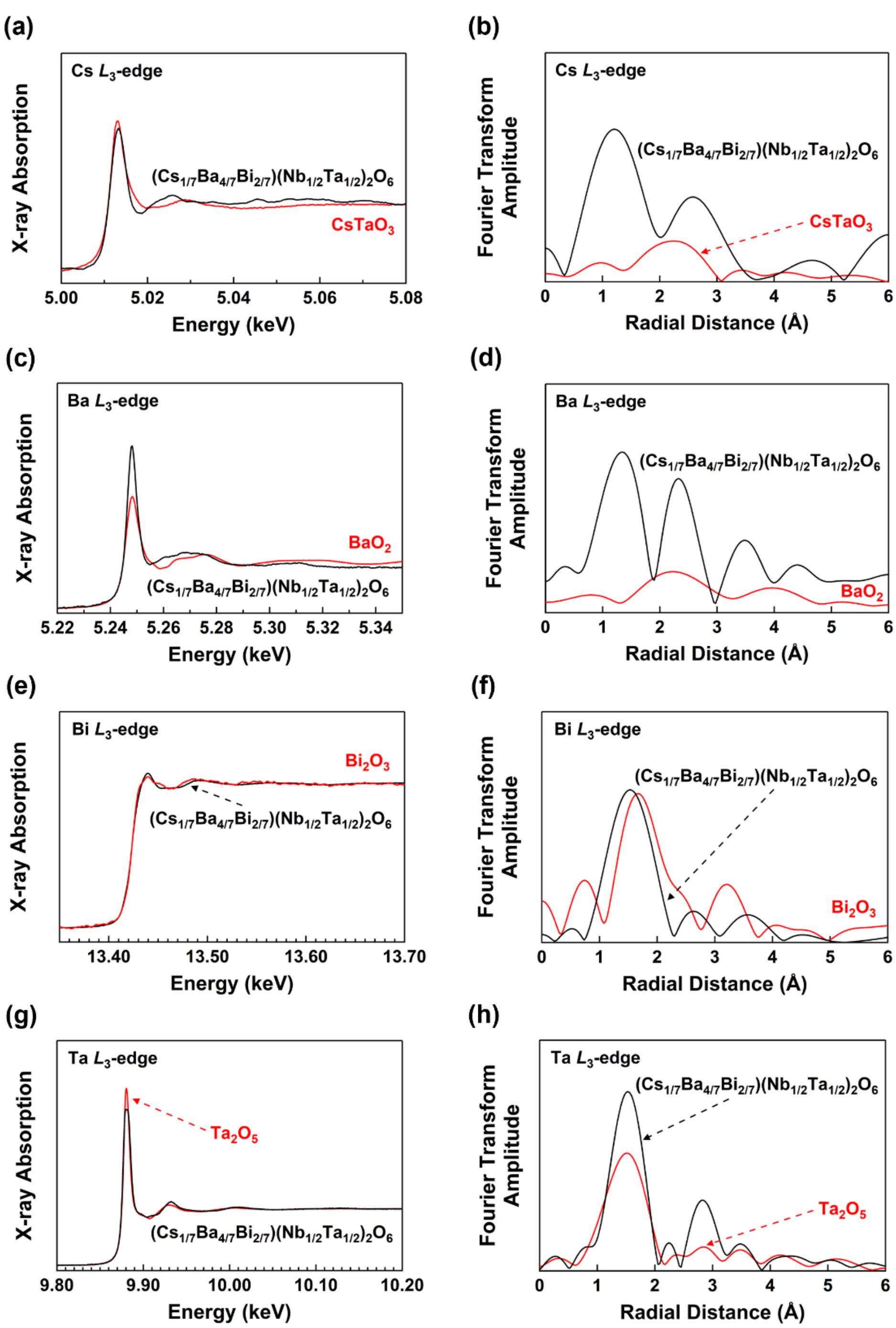

Fig. 6. The transformation in local electronic structure and neighbourhood of cations in HEO compared with relevant binary and ternary oxides. (a, c, e, g) XANES spectra and (b, d, f, h) Fourier-transformed EXAFS spectra in R-space for (a, b) Cs $L_3$-edge, (c, d) Ba $L_3$-edge, (e, f) Bi $L_3$-edge, (g, h) Ta $L_3$-edge of HEO and relevant binary and ternary oxides.

**Optical and spin characterizations**

Fig. 7a depicts the UV-Vis absorption spectrum of the HEO. A strong absorption edge is observed at a wavelength of 380 nm, indicating that the HEO absorbs the light mainly in the UV region. Through UV-Vis analysis by the Kubelka-Munk method, Fig. 7b shows the direct bandgap value of HEO at 3.0 eV. Fig. 7c depicts the UPS spectrum to estimate the valence band maximum (VBM) value. The measured VBM and secondary energy cut-off (SECO) values are 4.0 and 18.5 eV versus the Fermi level, respectively. The VBM versus vacuum was obtained based on Eq. 1, where $\phi$ denotes the work function, h corresponds to Planck's constant, $\upsilon$ denotes the photon frequency, and h$\upsilon$ is 21.2 eV. The band structure of HEO is described in Fig. 7d, where the VBM value is -6.7 eV versus vacuum or 2.26 eV versus normal hydrogen electrode (NHE). The conduction band minimum (CBM) value is obtained by subtracting the bandgap from the VBM, reaching -0.74 eV versus NHE. Compared with the band structures of $Nb_2O_5$ [39] (bandgap: 3.4 eV, VBM: 2.5 eV versus NHE, CBM: −0.9 eV versus NHE) and $Ta_2O_5$ [40] (bandgap: 3.9 eV, VBM: 3.49 eV versus NHE, CBM: −0.41 eV versus NHE), the HEO exhibits a narrower bandgap along with a modified band structure. This band structure of HEO satisfies the energy requirements for the $CO_2$ conversion process, i.e., the CBM of the HEO is higher than the reduction potential of $CO_2$/CO (-0.11 eV versus NHE) and $CO_2/CH_4$ (0.17 eV versus NHE).[11]

$$E_{\text{VBM vs. vacuum}} = -(E_{\text{VBM vs. Fermi level}} + h\upsilon - E_{\text{SECO}}) \quad (1)$$

Fig. 7e provides information on vacancy defects through ESR analysis. The signal with $g$ = 1.936 most likely originates from paramagnetic defects based on the Nb antisite.[41] Additionally, a small dual peak with a turning point at $g$ = 2.001 is characteristic of oxygen vacancies with an unpaired electron.[4] The $g$ value at 2.008 with high peak intensity is generally attributed to paramagnetic oxygen vacancies associated with lattice strain.[4] This result also supports the information on the XPS of O 1s as illustrated in Fig. 4f. The confirmation of the presence of oxygen vacancy defect sites also forms the basis for catalyst selection to enhance the efficiency of photocatalytic reactions. Several studies have shown these sites can be active sites that provide $CO_2$ uptake and generate more electrons to activate the $CO_2{\bullet}^-$ intermediate.[16,42] In addition, the PL spectrum in Fig. 7f also reveals the presence of a low-intensity peak at 600 nm, which shows that this HEO has small charge recombination. The intensity of this peak is about 40 times lower than the PL intensity for anatase $TiO_2$, examined under similar conditions. Improvements in the optical properties of the material are expected to contribute to improved performance of photocatalytic processes, such as the conversion of $CO_2$ to form more $CH_4$ and CO.

**Photocatalytic activity**

Fig. 8a, b, and c illustrate the photocatalytic $CO_2$ conversion process under four different conditions for the evolution of $H_2$, CO, and $CH_4$, respectively. In Experiment 1, only HEO was used without scavengers. Experiments 2, 3, and 4 were conducted under the same conditions as Experiment 1, with the introduction of PET micro-powder, microalgae, and methanol as sacrificial agents. All experiments were initially carried out for 4 h to reach a steady state, then continuous sampling was conducted every hour for 20 h. It can be observed that during $CO_2$ conversion using only the HEO, the average CO production reaches 5.6 µmol/g·h, while $CH_4$ evolution is approximately 1.1 µmol/g·h and no $H_2$ is detected. Based on the methane production rate and the CO formation rate, as well as the number of electrons required to reduce $CO_2$ into these products, the methane selectivity of HEO is 44%, while its CO selectivity is 56%.

$$CH_4 \text{ Selectivity} = 100 \times (8r_{CH4}) / (2r_{CO} + 8r_{CH4}) \quad (2)$$

Upon the addition of PET, the conversion efficiency increases, with CO and $CH_4$ concentrations reaching 25.9 and 1.6 µmol/g·h, respectively. The introduction of PET as a sacrificial agent reduces the recombination of electrons and holes, allowing electrons sufficient time to get involved in the $CO_2$ reduction process, contributing to the formation of more conversion products. When methanol is used as a conventional sacrificial scavenger, the production rate of CO and $CH_4$ does not increase and remains at 2.7 and 0.6 µmol/g·h, respectively. However, the $H_2$ production rate significantly increases with the addition of methanol and reaches 24.8 µmol/g·h. Microalgae also have a similar mechanism of action to PET and methanol as a scavenger for the reaction; however, the use of microalgae is more effective than PET and methanol. Specifically, in the presence of microalgae during $CO_2$ conversion by the HEO, the average evolution rates of CO and $CH_4$ increase to large values of 63.9 and 4.6 µmol/g·h, respectively. Additionally, $H_2$ is also produced at a rate of 37.0 µmol/g·h by the addition of microalgae, which is higher than that using methanol. To further evaluate the role of microalgae, two control experiments were conducted in the absence of the HEO, in which the microalgae were irradiated to assess their self-degradation capability. The first experiment was performed with a continuous stream of $CO_2$ gas, while the second one was conducted with a stream of argon gas. As shown in Fig. 8d, the results of two experiments show that after 24 h of irradiation, the production rate of CO and $CH_4$ reached approximately 20.0 and 2.0 µmol/g·h, respectively. These values are less than one-third of those obtained in the presence of HEO and microalgae, indicating that microalgae themselves can undergo degradation under strong irradiation to form products. However, in the presence of the photocatalyst, the efficiency is significantly enhanced. The stability of the HEO was evaluated through cycling tests, as shown in Fig. 8e. Moderate variations in product yields are observed without a clear trend. While the second cycle resulted in a slight increase in product concentration, the third cycle shows a slight decrease. These small changes may stem from differences in the shapes and sizes of the microalgal clusters, as shown in Fig. S1. Such structural heterogeneity can influence degradation rates, light absorption, mass transfer, and contact efficiency with the HEO catalyst, thereby leading to fluctuations in product yield during repeated cycles.

Fig. 8f characterizes the microalgae after photocatalysis through FT-IR analysis. The FT-IR spectra confirm the structure of *Chlamydomonas reinhardtii* microalgae, with major components including carbohydrates, lipids, and proteins. After photocatalysis, the peaks of the carbohydrate group have almost completely disappeared, while the peaks of the lipids and proteins have very low intensity. Based on the complexity of the structure of the lipids and proteins, their complete decomposition requires more time than the carbohydrate group. A comparison of XRD spectra (Fig. 8g) of the HEO before and after experiments confirms the long-term stability of the HEO. These results show that the use of HEO helps maintain the structural stability and photocatalytic activity, while microalgae significantly enhance the efficiency of $CO_2$ conversion products.

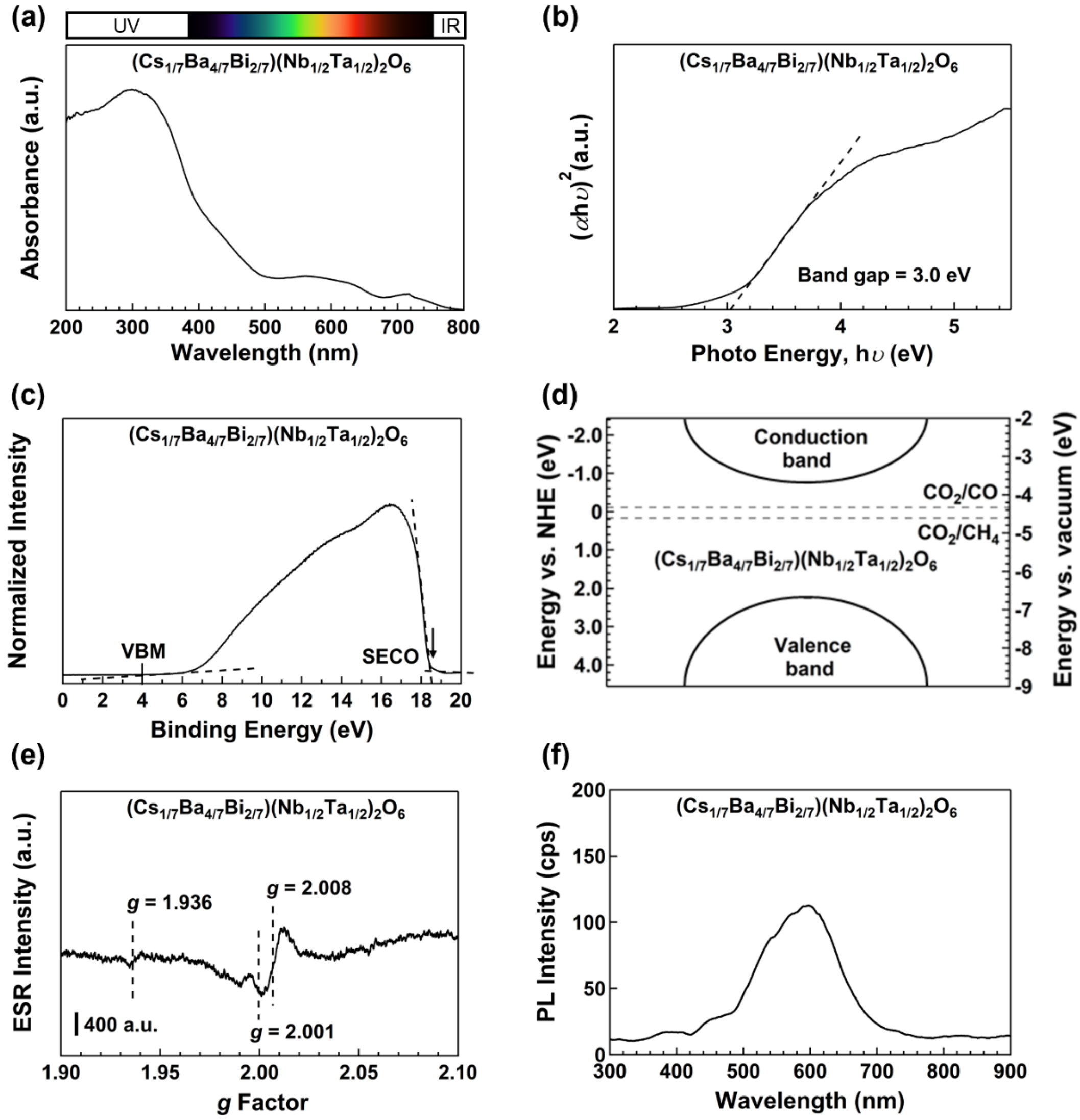

Fig. 7. Optical characterization and appropriate band structure of HEO for $CO_2$ conversion. (a) UV-Vis light absorbance analysis, (b) direct bandgap estimation based on Kulbelka-Munk method, (c) UPS analysis (He I, bias –4 V, VBM: valence band maximum, SECO: secondary electron cut-off), (d) band structure with valence band and conduction band values versus normal hydrogen electron and versus vacuum, and (e) ESR spectrum and (f) PL spectrum for HEO.

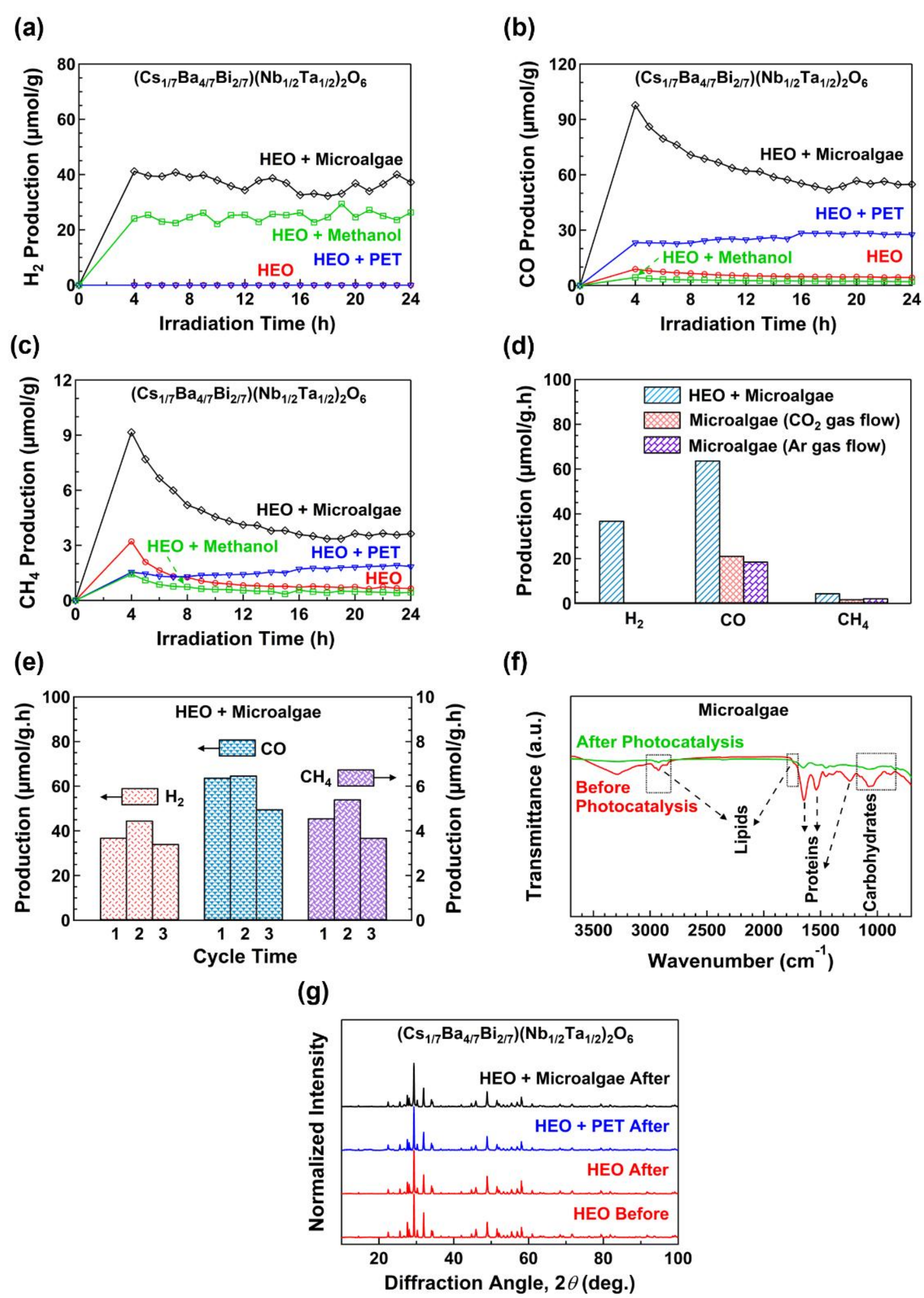

Fig. 8. Enhanced $CO_2$ conversion process efficiency using HEO as photocatalyst and microalgae as sacrificial agent. (a) $H_2$, (b) CO, and (c) $CH_4$ production from photocatalysis reactions under various conditions. (d) Average product concentration from photocatalysis and microalgae degradation. (e) Photocatalytic production rates for three cycles by using HEO and microalgae. (f) FT-IR spectra of microalgae before and after photocatalysis. (g) XRD analysis of HEO before and after $CO_2$ conversion.

## Discussion

Enhancing product yields from $CO_2$ photocatalytic conversion remains a major challenge in photocatalysis, because the strong C=O bonds make $CO_2$ highly stable during reduction reactions.[5] Research on high-entropy photocatalysts promoted innovation in this field by incorporating a variety of cations in a single crystal; however, the $CO_2$ conversion activity still needs further improvement. In this study, microalgae are used as hole scavengers for $CO_2$ conversion, contributing to improved yields of CO and $CH_4$. Besides, a novel HEO is designed and synthesized for this photocatalytic $CO_2$ conversion reaction. This section discusses in more detail two aspects: (i) evaluation of the performance of the HEO in comparison with other materials and reasons behind the outstanding catalytic activity of HEO, and (ii) the mechanism and the advantages of using microalgae as a scavenger in photocatalysis.

Regarding the first issue, as shown in Table 2, the HEO exhibits higher performance compared with conventional photocatalysts.[16,17,42-49] This improvement arises from several distinctive features related to material configuration. First, the HEO possesses an intriguing structural characteristic in which its two polymorphs share nearly identical frameworks: the cubic phase has a pyrochlore structure close to the one reported for $(A_xA'_y)B_2O_{6+z}$ ($1 < x + y < 1.3$, $z < 1$),[25,43,50] while the tetragonal phase resembles an $AB_2O_6$ layered perovskite or tungsten bronze structure.[26] The B-site atoms niobium and tantalum, with a coordination number of 6, form octahedral frameworks that connect at the corners of the lattice.[25,43] Several A sites with a coordination number of 8 create an elongated cube for the cubic phase. While other A sites with a coordination number of 12 form a perovskite-like structure, and others with a coordination number of 15 allow some octahedral rings to be inserted into the structure, forming a tetragonal phase.[26] The two phases share the same octahedral structure, reflecting their polymorphism in a single chemical composition. This is supported by SEM-EDS and STEM-EDS results (Fig. 3), which demonstrate the homogeneous distribution of all elements throughout the HEO. The significant similarity between the two phases leads to the formation of an electronic junction [3,51] while maintaining uniform elemental distribution, thereby enhancing charge separation and facilitating more efficient charge transport in the catalyst. Several previous studies have also reported the formation of dual-phase HEOs consisting of fluorite and pyrochlore structures within the same composition,[52] as well as order–disorder transitions between these structures.[53] In addition, phase stability (e.g., cubic versus tetragonal) is strongly influenced by factors such as cation size, disorder, and electronegativity differences,[54] further supporting this transition behavior. Besides, it should also be noted that the current authors observed that this polymorphic transition depends on the synthesis and post-synthesis conditions. The phase fractions used in this study are observed when the HEO is synthesized at 1373 K and rapidly cooled under atmospheric pressure. However, as demonstrated in Fig. S2, when the synthesized HEO is annealed at 673 K for 1 h, the fraction of the cubic phase increases, suggesting the occurrence of a tetragonal-to-cubic phase transformation. Second, catalyst design is based on the integration of elements niobium and titanium as the octahedral framework, while the three other elements (bismuth, cesium, and barium) provide benefits for the reaction. Earlier studies showed that mixing elements with different cationic configurations can provide both electron-donating and electron-accepting sites [11] and have been shown to improve charge-transfer processes in HEOs. In addition, the incorporation of alkali and alkaline earth metals such as cesium and barium increases surface basicity, enhances $CO_2$ chemisorption, and suppresses charge recombination.[11,41,42] Moreover, the incorporation of bismuth in the form of $Bi^{3+}$ introduces a stereochemically active $6s^2$ lone pair.[55] This lone pair exerts a strong polar distortion on the local coordination environment, generating a localized dipole moment (i.e., localized polarization). This internal electric field actively drives the separation of photogenerated electron–hole pairs, suppressing their recombination.[55] Finally, the formation of oxygen vacancies, as evidenced by the O 1s XPS and ESR spectra, is considered to generate active sites for photocatalytic reactions.[3,32] Taken together, all these features synergistically contribute to the enhanced $CO_2$ reduction performance of the HEO.

Regarding the second issue, the results shown in Fig. 8 can be used to elucidate a $CO_2$ conversion mechanism in which microalgae act as a sacrificial agent. It is clearly observed that, in the presence of microalgae, the CO production efficiency increases by 10-fold and that of $CH_4$ by 4-fold compared with the photocatalytic process using only HEO. Microalgae as a sacrificial agent also demonstrates higher performance when compared with PET plastic, a common environmental pollutant with high potential for photoreforming,[35,36] used in this study as a sacrificial agent. This can be explained by the fact that microalgae are more degradable and can be readily decomposed compared to plastic under light irradiation in a neutral pH environment. As a result, smaller molecular components are generated, which can effectively scavenge photogenerated holes and promote charge separation. In addition, microalgae possess a complex composition leading to multiple different oxidation pathways, which contribute to enhanced overall reaction kinetics. After the photocatalysis, the microalgal clusters were almost completely degraded. FT-IR analyses indicate that easily degradable carbohydrate groups disappear, while only weak intensities corresponding to the characteristic peaks of lipid and protein groups remain. Additionally, GC–MS analysis was conducted, as described in Fig. S3 of the supplementary information; however, no specific components were detected in the liquid phase extracted after the reaction. The decomposition of microalgae contributes to enhanced $CO_2$ reduction and the formation of a greater quantity of products such as CO and $CH_4$. As schematically shown in Fig. 9, the use of microalgae as a sacrificial agent not only diminishes the necessity of using alcohols whose production emits $CO_2$ but also contributes to $CO_2$ capture because microalgae capture $CO_2$ during their cultivation through the photosynthesis process. [22,23]

It should be noted that the morphology and size of the microalgae clusters can affect the efficiency of the photocatalytic process. Further studies could evaluate the effects of different microalgal strains and optimize their dosage as well as particle size. Future experiments on ternary or quaternary $AB_2O_6$-type oxides by omitting or substituting individual elements can also further clarify the effect of individual elements in the HEO on the photocatalytic $CO_2$ conversion mechanism in the presence of microalgae. A detailed life cycle assessment and carbon balance estimation, together with electrochemical analyses, should also be applied to fully evaluate the environmental and charge transfer impact of photocatalysis. Finally, the design of high-entropy photocatalysts that can function under visible light and ultimately sunlight can be an important step towards realizing energy efficiency in the microalgae-based photocatalytic system. This study, together with an earlier study,[56] suggests the high potential of bio-scavengers in photocatalysis, although optimization of the process and catalyst should be studied in the future.

Table 2. Summary of photocatalytic CO and CH4 production using different catalysts

| Catalyst | Reaction system | Scavenger for Holes | Co-catalyst | $CO_2$ conversion (µmol/g.h) | | | Ref. |
|---|---|---|---|---|---|---|---|
| | | | | CO | $CH_4$ | $H_2$ | |
| HEO (0.1 g) | - Continuous gas-solid reactor | - | - | 5.6 | 1.1 | - | This study |
| | - Light: 400 W mercury | - Methanol (50 mL) | - | 2.7 | 0.6 | 24.8 | |
| | | - PET plastic (0.1 g) | - | 25.9 | 1.6 | - | |
| | | - Microalgae (0.1 g) | - | 63.9 | 4.6 | 37.0 | |
| $ZnTa_2O_6$ (0.5 g) | - Continuous gas-solid reactor<br>- Light: 400 W mercury | - | 3 wt% Ag | 43.7 | - | - | [16] |
| $SrNb_2O_6$ (0.01 g) | - Continuous gas-solid reactor<br>- Light: 100 W xenon | - | - | 16.6 | 3.3 | 6.5 | [17] |
| $In_2O_3$ with oxygen vacancy (0.01 g) | - Continuous gas-solid reactor<br>- Light: 300 W xenon | Triethanolamine (2 mL) | - | 63.3 | 0.8 | 11.6 | [42] |
| $CsBi_2Ta_5O_{16}$ | - Batch gas-solid reactor | - | - | 1.1 | 0.2 | - | [43] |
| (0.1 g) | - Light: 300 W xenon | - | 10 wt% NiO | - | 7.2 | - | |
| $NaNbO_3$ | - Continuous gas-solid reactor | - | - | 6.47 | 0.015 | 2.9 | [44] |
| $NaTaO_3$ | - Light: 4 fluorescents, a total | - | - | 0.06 | 0.006 | 0.12 | |
| $NaNb_{0.5}Ta_{0.5}O_3$ (0.1 g) | irradiance of 71.7 $Wm^{-2}$ | - | - | 2.4 | 0.03 | 2.4 | |
| P25 $TiO_2$ (0.2 g) | - Batch gas-solid reactor<br>- Light: 300 W xenon | Methanol (20 mL) | - | 1.3 | - | 2.7 | [45] |
| Porous $Ta_2O_5$ | - Continuous gas-solid reactor | - | - | 0.3 | - | 1.5 | [46] |
| (0.1 g) | - Light: 300 W xenon | - | 1 % Ag | 1.8 | - | 0.2 | |
| $SnTa_2O_6$ | - Continuous gas-solid reactor | - | - | 28 | - | - | [47] |
| $Sn_2Ta_2O_7$ (0.02 g) | - Light: 350 W xenon | - | - | 1.7 | - | - | |
| $Bi_4O_5Br_2$ (0.02 g) | - Batch gas-solid reactor<br>- Light: 300 W xenon | - | - | 31.6 | - | - | [48] |
| $ZnIn_2S_4$ (0.05 g) | - Batch gas-solid reactor | Triethanolamine | Undoped | 1.6 | 0.03 | 4.1 | [49] |
| | - Light: 300 W xenon (λ ≥ 420 nm) | (10 mL) | Er-doped | 0.6 | 6.7 | 1.6 | |

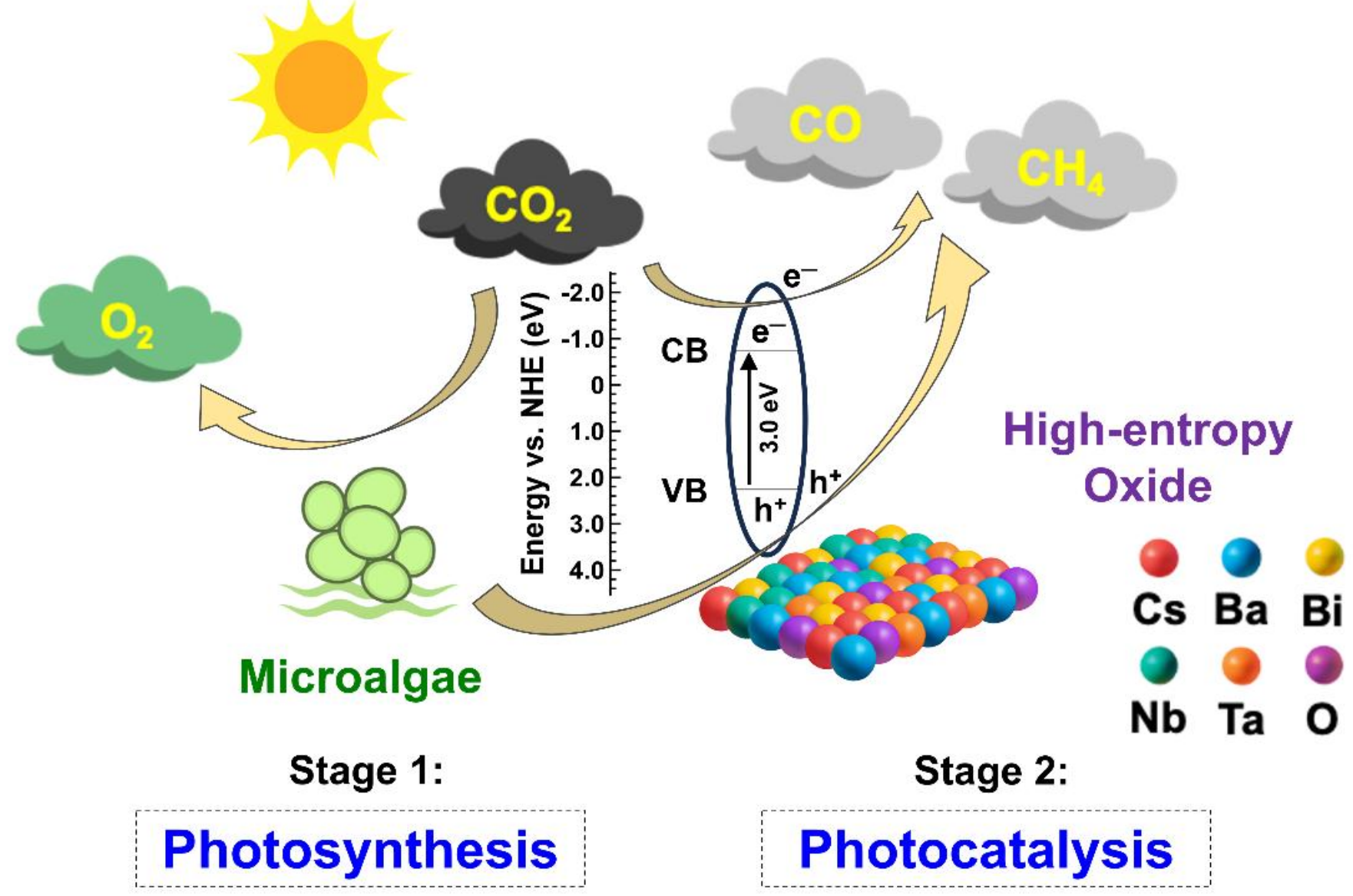

Fig. 9. Schematic illustration of photocatalytic $CO_2$ conversion using microalgae as hole sacrificial agent, realized by HEO.

## Conclusion

In this study, a novel pathway for $CO_2$ conversion is suggested by using microalgae as a sacrificial agent in photocatalysis. For this pathway, an $AB_2O_6$-type high-entropy material, $(Cs_{1/7}Ba_{4/7}Bi_{2/7})(Nb_{1/2}Ta_{1/2})_2O_6$, is synthesized, which has bi-polymorphism (layered perovskite and pyrochlore) while maintaining high homogeneity of all elements. The new pathway exhibits an improvement in the production efficiency of $CH_4$ and CO. These results indicate that the use of microalgae is beneficial not only for $CO_2$ capture during its cultivation but also for $CO_2$ conversion during photocatalysis, suggesting a sustainable route for the carbon cycle.

## Author contributions

All authors participated in conceptualization, investigation, methodology, validation, and writing - review & editing.

## Conflicts of interest

The authors declare no conflict of interest.

## Data availability

Data will be made available on request.

## Acknowledgements

H.T.N.H. would like to express gratitude to the Yoshida Scholarship Foundation (YSF) for its support in the form of a Ph.D. scholarship. This research was funded in part by Mitsui Chemicals, Inc., Japan; in part by the ASPIRE initiative of the Japan Science and Technology Agency (JST) (JPMJAP2332), and in part by the São Paulo Research Foundation (FAPESP), grant number 2024/01639-0.

## Notes and references